\documentclass[11pt,a4paper]{article}
\usepackage{jheppub}

\usepackage{amsmath, amsfonts, amsthm, amssymb, graphicx}
\usepackage[dvipsnames]{xcolor}
\usepackage{epstopdf}
 \usepackage{slashed}
\usepackage{changepage}
 \usepackage{vector}

\usepackage{pifont}% http://ctan.org/pkg/pifont

\usepackage{caption}
\def\be{\begin{equation}}
\def\ee{\end{equation}}
\def\ba{\begin{eqnarray}}
\def\ea{\end{eqnarray}}

\def\bdm{\begin{displaymath}}
\def\edm{\end{displaymath}}

\def\bq{\begin{quote}}
\def\eq{\end{quote}}

\def\ltap{\ \raise.3ex\hbox{$<$\kern-.75em\lower1ex\hbox{$\sim$}}\ }
\def\gtap{\ \raise.3ex\hbox{$>$\kern-.75em\lower1ex\hbox{$\sim$}}\ }
\def\gl{\ \raise.5ex\hbox{$>$}\kern-.8em\lower.5ex\hbox{$<$}\ }
\def\roughly#1{\raise.3ex\hbox{$#1$\kern-.75em\lower1ex\hbox{$\sim$}}}

 at 10truept

\newcommand{\beq}{\begin{equation}}
\newcommand{\eeq}{\end{equation}}
\newcommand{\bea}{\begin{eqnarray}}
\newcommand{\eea}{\end{eqnarray}}
\newcommand{\beqa}{\begin{eqnarray}}
\newcommand{\eeqa}{\end{eqnarray}}

\def\be{\begin{equation}}
\def\ee{\end{equation}}

\usepackage[dvipsnames]{xcolor}

\title{Observational Tests of 4D Double Field Theory}

\author[1]{Shunrui Li,}  
\author[2,3]{Yang Liu}

\affiliation[1]{Department of mathematicas and applied mathematics, Beijing University of Chemical Technology, Beijing 100029, China}
\affiliation[2]{School of Physics and Astronomy, University of Nottingham, University Park, Nottingham NG7 2RD, United Kingdom}
\affiliation[3]{Nottingham Centre of Gravity, University of Nottingham, Nottingham NG7 2RD, UK}

\emailAdd{shunruili687@gmail.com}
\emailAdd{yang.liu@nottingham.ac.uk}

\abstract{Although General Relativity (GR) is a very successful theory of gravity, it cannot explain every observational phenomenon. People have tried many kinds of modified gravity theory to explain these phenomena which GR cannot explain very well, such as string theory. In recent years Double Field Theory (DFT) has been an exciting research area in string theory. The most general, spherically symmetric, asymptotically flat, static vacuum solution to $D=4$ double field theory has been derived by S.M. Ko, J.H. Park and M. Suh. In this article, we calculate the minor corrections to the three predictions in GR: optical deflation, planet precession and gravitational redshift. These three predictions should be able to tested by observations and find the discrepancies between GR and DFT in the future.  } 

\begin{document}
\maketitle

\section{Introduction}
At present General Relativity (GR) is the most successful theory of gravity, which has been verified by observations accurately. Although GR is very successful, it still cannot explain every observational phenomenon. Take the galaxy rotation curve as an example. While GR predicts the Keplerian inverse square root fall-off of the velocities, however, observations show rather ‘flat’ ($\sim 150$ km/s) or ‘rising’ curves \cite{KPS2017, RTF1980}. The resolution of the discrepancy might call for dark matter or modifications of the theory of gravity \cite{KPS2017,M1983}, or perhaps both. So far the most promising quantum gravity theory is string theory, which is believed to unify four fundamental interactions. Therefore, it is natural to apply string theory to explain the phenomena which GR cannot explain successfully, such as the galaxy rotation curve. \\
On the one hand, in general relativity the metric, $g_{\mu\nu}$, is the only geometric object. All other fields, which source the gravity, are viewed as matter or radiation. On the other hand, string theory puts a two-form gauge potential and a scalar dilaton on an equal footing along with the metric, since the three of them, conventionally denoted by $B_{\mu\nu}$, $\phi$, $g_{\mu\nu}$,  correspond to the massless NS-NS sector and form a multiplet of T-duality \cite{KPS2017}. This may indicate the existence of an alternative theory of gravity where the whole massless NS-NS sector becomes geometric as the gravitational unity. Such an idea has been realized in recent years through the developments of Double Field Theory (DFT) \cite{KPS2017}.\\
In this article, we intend to consider the observational tests of 4d Double Field Theory (DFT), which is an exciting research area in string theory in recent years. The primary goal of DFT \cite{KPS2017, Siegel19931, Siegel19932, HZ20091} was to reformulate supergravity with doubled coordinates, namely, $x_A = (\tilde{x}_{\mu}, x^{\nu})$, in a way that T-duality has been a manifest symmetry of the action and unifies diffeomorphism and B-field gauge symmetry into ‘doubled diffeomorphism’ \cite{HZ20092,HZ20101,HZ20102}. The most general, spherically symmetric, asymptotically flat, static vacuum solution to $D=4$ double field theory has been derived directly in \cite{KPS2017}. The equivalent form of the solution can be obtained by using S-duality \cite{BMQ1994}. Furthermore, Stephen Angus, Kyoungho Cho and Jeong-Hyuck Park studied more properties of Einstein double field equations in \cite{ACP2018}. In \cite{CP2022}, Choi and Park performed post-Newtonian analysis of double field theory as a test of string theory in gravitational sector against observations. In \cite{LPVY2023} the authors claimed that 4D DFT has passed a test for late-time cosmology. Yang Liu analyzed Hawking radiation of the solution \cite{YL2022}, which was obtained in \cite{KPS2017}.  \\
In our article, we intend to study some observational tests of the solution obtained in ref.\cite{KPS2017}. The article is composed as follows: in section 2, we review the double field theory and spherical solutions in 4D double field theory briefly; in section 3, we obtain three observational tests for the spherical solution; in section 4, the results of the article have been discussed.

\section{Review of 4D Double Field Theory}
\subsection{Review of double field theory}
Double field theory (DFT) generalizes the spacetime action, which can possess T-duality on the level of component fields manifestly \cite{WY2014}. Earlier efforts can be traced back to \cite{T1990,D1990}. Due to the equivalence of spacetime momenta and winding numbers in the string spectra \cite{WY2014}, a set of conjugated coordinates $\tilde{x}^i$, which is conjugated to winding numbers, can be introduced naturally \cite{WY2014}. These conjugated coordinates are treated on the same footing as the usual coordinates $x_i$. Then the dimension of spacetime is doubled from $D$ to $D+D$ \cite{WY2014}. \\
The action of DFT unifies the metric $g_{ij}$, the two-form $b_{ij}$ and the dilaton $\phi$ by rewriting these fields in an $O(D,D)$ covariant way. If there is no dependence on the conjugated coordinates, the DFT can be reduced to the supergravity action \cite{B2019}. The action is given by 
\begin{equation}\label{eq:2.1}
S = \int dx d\tilde{x} e^{-2d} \mathcal{R}, 
\end{equation}
where $d$ is the shifted dilaton, which contains the determinant of the metric and the usual dilaton $\phi$ both \cite{B2019,HHZ2010}, i.e.,
\begin{equation}\label{eq:2.2}
e^{-2d} = \sqrt{-g} e^{-2\phi}, 
\end{equation}
and 
\begin{equation}\label{eq:2.3}
\mathcal{R}= \frac{1}{8} \mathcal{H}^{MN} \partial_{M} \mathcal{H}^{KL} \partial_{N} \mathcal{H}_{KL} - \frac{1}{2} \mathcal{H}^{MN} \partial_{N} \mathcal{H}^{KL} \partial_{L} \mathcal{H}_{MK} - \partial_{M} d \partial_{N} \mathcal{H}^{MN} + 4 \mathcal{H}^{MN} \partial_{M} d  \partial_{N} d,
\end{equation}
where the generalized metric $\mathcal{H}_{MN}$ is defined as 
\begin{equation}\label{eq:2.4}
\mathcal{H}_{MN}
=
\begin{bmatrix}
g^{ij} & -g^{ik} b_{kj}\\
b_{ik}g^{kj} & g_{ij}- b_{ik} g^{kl} b_{lj}
\end{bmatrix}.
\end{equation}
The level matching condition in closed string theory imposes the “weak constraint”, $\partial \tilde{\partial} \phi (x, \tilde{x}) = 0$, for any field $\phi (x, \tilde{x})$. Furthermore, in order to ensure the action locally equivalent to the low energy effective string action, the “strong constraint”, $\partial \tilde{\partial} = 0$, is required as an operator equation, acting on any products of the fields \cite{WY2014}. 

\subsection{Spherical solutions in D=4 double field theory}
In this section, we review the most general form of the static, asymptotically flat and spherically symmetric vacuum solutions to $D=4$ double field theory briefly \cite{KPS2017, ACP2018}. Without loss of generality, the metric for the string frame can be assumed as 
\begin{equation}\label{eq:2.5}
ds^2 = e^{2\phi(r)} [-A(r)dt^2 + A(r)^{-1} dr^2 + A(r)^{-1} C(r)d\Omega^2], 
\end{equation}
where
\begin{equation}\label{eq:2.6}
 d\Omega^2= d\theta^2 + \sin^2 \theta d\varphi^2.
\end{equation}
It is worthwhile to note that our string frame metric ansatz takes the product form of the dilaton factor, $e^{2\phi}$, times the Einstein frame metric \cite{KPS2017}.\\
If the spacetime is asymptotically 'flat', then the metric $(2.5)$ should satisfy the following three boundary conditions \cite{KPS2017}:
\begin{equation}\label{eq:2.7}
\lim_{r \rightarrow \infty} A(r) =1,
\end{equation} 
\begin{equation}\label{eq:2.8}
\lim_{r \rightarrow \infty} r^{-2} C(r) =1,
\end{equation}
\begin{equation}\label{eq:2.9}
\lim_{r \rightarrow \infty} \phi(r) =0.
\end{equation}
From the asymptotic ‘smoothness’, the metric $(2.5)$ should satisfy \cite{KPS2017}:
\begin{equation}\label{eq:2.10}
\lim_{r \rightarrow \infty} A'(r) = \lim_{r \rightarrow \infty} A''(r) = 0,
\end{equation} 
\begin{equation}\label{eq:2.11}
\lim_{r \rightarrow \infty} r^{-1} C'(r) = \lim_{r \rightarrow \infty} C''(r) =2,
\end{equation}
\begin{equation}\label{eq:2.12}
\lim_{r \rightarrow \infty} \phi'(r) = \lim_{r \rightarrow \infty} \phi''(r) =0.
\end{equation}
We write the $B$-field using the form notation \cite{KPS2017}
\begin{equation}\label{eq:2.13}
B_{(2)}= \frac{1}{2} B_{\mu\nu} dx^{\mu} \wedge dx^{\nu} = B(r) \cos \theta dr \wedge d\varphi + h \cos \theta dt \wedge d\varphi.
\end{equation}
The $H$-flux, which takes the most general spherically symmetric form, can be written as \cite{KPS2017}
\begin{equation}\label{eq:2.14}
H_{(3)}= \frac{1}{3!} H_{\lambda\mu\nu} dx^{\lambda} \wedge dx^{\mu} \wedge dx^{\nu} = B(r) \sin \theta dr \wedge d\theta \wedge d\varphi + h \sin \theta dt \wedge d\theta \wedge d\varphi,
\end{equation}
which is closed for constant $h$ \cite{KPS2017}. As a result, with four constants $a$, $b$, $c$, $h$ and \cite{KPS2017}
\begin{equation}\label{eq:2.15}
c_{+} = c + \frac{1}{2} \sqrt{a^2 + b^2},
\end{equation}
\begin{equation}\label{eq:2.16}
c_{-} = c - \frac{1}{2} \sqrt{a^2 + b^2},
\end{equation}
\begin{equation}\label{eq:2.17}
\gamma_{\pm} = \frac{1}{2} (1 \pm \sqrt{1-h^2/b^2}),
\end{equation}
the metric $(2.5)$ can be given \cite{KPS2017, ACP2018}:
\begin{equation}\label{eq:2.18}
e^{2\phi} = \gamma_{+} (\frac{r- \alpha}{r+\beta})^{\frac{b}{\sqrt{a^2 + b^2}}} +  \gamma_{-} (\frac{r+ \beta}{r-\alpha})^{\frac{b}{\sqrt{a^2 + b^2}}}, 
\end{equation}
\begin{equation}\label{eq:2.19}
B_{(2)} = h \cos \theta dt \wedge d\varphi, 
\end{equation}
\begin{equation}\label{eq:2.20}
H_{(3)} = h \sin \theta dt \wedge d\theta \wedge d\varphi, 
\end{equation}
and
\begin{equation}\label{eq:2.21}
ds^2= e^{2\phi} [- (\frac{r- \alpha}{r+\beta})^{\frac{a}{\sqrt{a^2 + b^2}}} dt^2 +  (\frac{r+ \beta}{r-\alpha})^{\frac{a}{\sqrt{a^2 + b^2}}} \times \{dr^2 + (r-\alpha)(r+\beta) d\Omega^2\} ],
\end{equation}
where
\begin{equation} \label{alpha}
\alpha=\frac{a}{a+b} \sqrt{a^2+b^2},
\end{equation}
and
\begin{equation} \label{beta}
\beta=\frac{b}{a+b} \sqrt{a^2+b^2}.
\end{equation}
If the metric must be real, then we must require $b^2 \geq h^2$ \cite{KPS2017}. Eqs.$(2.18)$-$(2.21)$ are the most general form of the static, asymptotically flat and spherically symmetric vacuum solutions to 4D double field theory \cite{KPS2017, ACP2018}. We should point out that although the fundamental symmetry principle of DFT is the backbone of the present work, in fact, with the ansatz $(2.5)$ and $(2.13)$, we are solving the full Euler-Langrangian equations of the familiar NS-NS sector gravity, namely, \cite{KPS2017}
\begin{equation}\label{eq:2.22}
S =\int d^4 x \sqrt{-g} e^{-2\phi} (R_g + 4\partial_{\mu} \phi \partial^{\mu} \phi - \frac{1}{12} H_{\mu\nu\rho} H^{\mu\nu\rho}),
\end{equation}
as they are equivalent to the vanishing of the both two indexed and zero-indexed DFT-curvatures, which is DFT vacuum. It can be proved that the asymptotic flatness is inconsistent with the magnetic $H$-flux, therefore we set $B(r)=0$ and $H_{(3)} = h\sin \theta dt \wedge d\theta \wedge d\varphi$ \cite{KPS2017}.
We can define “proper” radius:
\begin{equation}\label{eq:2.23}
R \equiv \sqrt{g_{\theta \theta} (r)} = \sqrt{C(r)/A(r)} e^{\phi(r)},
\end{equation}
then the angular part of the metric can be properly normalized \cite{KPS2017}:
\begin{equation}\label{eq:2.24}
ds^2 = g_{tt} dt^2 + g_{RR}dR^2 + R^2 d\Omega^2 = - e^{2\phi}A dt^2 + e^{2\phi} A^{-1} \left(\frac{dR}{dr} \right)^{-2} dR^2 + R^2 d\Omega^2. 
\end{equation}
If $b=h=0$ and $a=2M_{\infty}G > 0$, the Schwarzschild metric will be recovered: with proper radius, we have
\begin{equation}\label{eq:2.25}
ds^2 = -\left(1- \frac{2M_{\infty}G}{R} \right)dt^2 + \left(1- \frac{2M_{\infty}G}{R} \right)^{-1} dR^2 + R^2d\Omega^2, \qquad \phi=0, \qquad B_{\mu\nu}=0.
\end{equation}
If $b=h=0$ and $a=2M_{\infty}G < 0$, then
\begin{equation}\label{eq:2.26}
ds^2 = -\left(1+ \frac{2M_{\infty}G}{r} \right)^{-1}dt^2 + \left(1+ \frac{2M_{\infty}G}{r} \right) dr^2 + (r+2M_{\infty}G )^2 d\Omega^2, \qquad \phi=0, \qquad B_{\mu\nu}=0.
\end{equation}
After the radial coordinate redefinition, $r \rightarrow R-2M_{\infty}G $, the metric $(2.26)$ will reduce to $(2.25)$ with negative mass \cite{KPS2017}.\\

\section{Three observational tests of 4D Double Field Theory}
In this section, we consider the observational tests of $D=4$ Double Field Theory. Albert Einstein proposed three tests for General Relativity (GR): optical deflation, Mercury precession and gravitational redshift \cite{WeinbergGC}. In the following we will calculate the minor corrections to the three predictions in GR. In section 2, we have known that if $b=h=0$ and $a=2M_{\infty}G > 0$, the Schwarzschild metric is recovered. Therefore, assuming $b$ is small and $h=0$ (since the electric $H$-flux must be zero due to an energy condition \cite{CP2022}), and $a=2M_{\infty}G > 0$, in this section we will investigate the minor corrections to the three predictions in GR \footnote{In fact, in 4D DFT, we can obtain that $GM= \frac{1}{2}(a + b \sqrt{1-h^2/b^2}) = \frac{1}{2} (a+b)$, where $M$ is the Newtonian mass and $h=0$ \cite{CP2022}. In order to compare the theoretical results with observational data, it is convenient to take $a=2GM_{\infty}$, where $M_{\infty}$ can be obtained from observational data. And we regard $b$ as the minor correction to GR. The PPN parameters, $\beta_{PPN}$ and $\gamma_{PPN}$, are widely used in the literature of modified gravity \cite{CP2022}. But in order to compare the results with the standard results in GR directly, we will not use PPN parameters in this article. }. Since $b\ll a$, we just consider the first terms in an expansion in powers of $\frac{b}{a}$. All the following calculations are done in the Einstein frame, since all the calculations in GR are also done in the Einstein frame, which is easier to compare the theoretical results with the observational data. Therefore, according to \eqref{eq:2.21}, the metric, which we will consider, is the following:
\begin{equation}\label{Emetric}
ds^2= - (\frac{r- \alpha}{r+\beta})^{\frac{a}{\sqrt{a^2 + b^2}}} dt^2 +  (\frac{r+ \beta}{r-\alpha})^{\frac{a}{\sqrt{a^2 + b^2}}} \times \{dr^2 + (r-\alpha)(r+\beta) d\Omega^2\} ,
\end{equation}
where
\begin{equation} \label{alpha2}
\alpha=\frac{a}{a+b} \sqrt{a^2+b^2},
\end{equation}
\begin{equation} \label{beta2}
\beta=\frac{b}{a+b} \sqrt{a^2+b^2}.
\end{equation}
and
\begin{equation}\label{eq:dOmega2}
 d\Omega^2= d\theta^2 + \sin^2 \theta d\varphi^2.
\end{equation}

\subsection{Optical deflation in 4D DFT}
Since the rest mass of photon is zero, then we introduce an affine parameter $\lambda$ and define 4-momentum of photon:
\begin{eqnarray} \label{311}
p^{\mu} = \frac{dx^{\mu}}{d\lambda}.
\end{eqnarray}
For photon, we have
\begin{eqnarray} \label{312}
g_{\mu \nu } p^{\mu } p^{\nu  }  =  0,
\end{eqnarray}
namely,
\begin{eqnarray}\label{313}
g_{00} \left( \frac{dt}{d\lambda } \right)^2+ g_{11} \left(\frac{dr}{d\lambda } \right)^2+g_{22} \left(\frac{d\theta }{d\lambda } \right)^2+g_{33} \left(\frac{d\varphi }{d\lambda } \right)^2 & = & 0.
\end{eqnarray}
%\begin{eqnarray} \label{313}
%p_{0} & = & g_{0\nu } p^{\nu } 
%\end{eqnarray}
Since the metric is isotropic, it can be considered that the orbit of photon is limited to the equatorial plane, namely, $\theta = \frac{\pi}{2}$, therefore we have $\frac{d\theta}{d\lambda}=0$.\\ 
For the metric \eqref{Emetric}, $p_0$ and $p_3$ are conserved, we can get
\begin{eqnarray} \label{314}
g_{00}\frac{dt}{d\lambda }  = - E,
\end{eqnarray}
and
\begin{eqnarray} \label{315}
g_{33}\frac{d\varphi }{d\lambda } =  L,
\end{eqnarray}
where $E$ and $L$ represent energy and angular momentum of unit mass, respectively. \\
According to \eqref{311}-\eqref{315}, we can have
\begin{eqnarray} \label{316}
\left(\frac{dr}{d\lambda }\right)^2 = \frac{-g_{33}(\frac{d\varphi }{d\lambda } )^2-g_{00}(\frac{dt }{d\lambda } )^2}{g_{11}}= \frac{-\frac{L^2}{g_{33}} -\frac{E^2}{g_{00}}}{g_{11}}, 
\end{eqnarray}
\begin{eqnarray} \label{317}
\left(\frac{d\varphi }{dr} \right)^2 = \frac{(\frac{d\varphi }{d\lambda })^2}{(\frac{dr }{d\lambda })^2} = \frac{\frac{g_{11}}{g_{33}}L^2 }{-L^2-\frac{g_{33}}{g_{00}}E^2 }. 
\end{eqnarray}
According to \eqref{Emetric}, we have
\begin{eqnarray} \label{318}
\frac{g_{11}}{g_{33}} = \frac{1}{(r-\alpha)(r+\beta)} ,
\end{eqnarray}
\begin{eqnarray} \label{319}
\frac{g_{33}}{g_{00}}= -\frac{(r-\alpha)(r+\beta)}{\left(\frac{r-\alpha}{r+\beta} \right)^{\frac{2a}{\sqrt{a^2+b^2}}}}.
\end{eqnarray}
As a result, if we take $a= 2G M_{\infty}$, we can obtain the following equation around $b=0$:
\begin{eqnarray} \label{320}
\left(\frac{1}{r^2} \frac{dr}{d\varphi}\right)^2 = \left(\frac{E}{L}\right)^2 -\frac{1}{r^2} \left(1- \frac{2GM_{\infty}}{r} \right) + \frac{4}{r} \left(\frac{E}{L}\right)^2 b - \frac{1}{r^3} \left(2- \frac{2GM_{\infty}}{r} \right) b.
\end{eqnarray}
The first two terms on the right hand side of \eqref{320} are the contributions from General Relativity, while the other two terms on the right hand side of \eqref{320} are the contributions from 4D Double Field Theory.\\
If we define $\mu = \frac{G M_{\infty}}{r}$, then from \eqref{320} we have
\begin{eqnarray} \label{321}
\frac{d^2 \mu}{ d\varphi^2} + \mu = 3\mu^2 - \frac{3 \mu^2}{G M_{\infty}} b + 2GM_{\infty} \left( \frac{E}{L}\right)^2 b + \frac{4 \mu^3}{G M_{\infty}} b. 
\end{eqnarray}
We only consider the case where $G M_{\infty}$ is always much smaller than $r$. And considering that $b$ is very small, then the last term on the right hand side of \eqref{321} can be neglected, namely,
\begin{eqnarray} \label{322}
\frac{d^2 \mu}{ d\varphi^2} + \mu \approx 3\mu^2 - \frac{3 \mu^2}{G M_{\infty}} b + 2GM_{\infty} \left( \frac{E}{L}\right)^2 b. 
\end{eqnarray}
An approximate special solution of \eqref{322} is
\begin{eqnarray} \label{323}
\begin{aligned} 
    \mu & = \mu_0 \cos{\varphi} + \mu^2_0 (1+ \sin^2{\varphi}) \\ 
         & + \frac{1}{12} (3A\cos^2{\varphi} + 12B \cos^2{\varphi} + A \cos{\varphi} \cos{3 \varphi} + 9 A \sin^2{\varphi} + 12 B \sin^2{\varphi} + A \sin{\varphi} \sin{3\varphi} ), \\
\end{aligned}
\end{eqnarray}
where 
\begin{eqnarray} \label{324}
A = -\frac{3\mu^2_0}{G M_{\infty}} b,
\end{eqnarray}
\begin{eqnarray} \label{325}
B = 2 G M_{\infty} \left( \frac{E}{L}\right)^2 b,
\end{eqnarray}
\begin{eqnarray} \label{326}
\mu_0 = \frac{G M_{\infty}}{R},
\end{eqnarray}
and $R$ is the radius of the star. On the right hand side of \eqref{323}, the first two terms are from General Relativity and the third term is the contribution from 4D DFT.\\
Considering the azimuth angle at $\pm (\frac{\pi}{2} + \alpha)$, where $\alpha$ is a small quantity, then from \eqref{323} we have
\begin{eqnarray} \label{327}
-\mu_0 \alpha + 2 \mu^2_0 - \frac{2 \mu^2_0}{G M_{\infty}} b + \frac{2 G M_{\infty}}{\mu_0} \left( \frac{E}{L} \right)^2 b \approx 0,
\end{eqnarray}
namely,
\begin{eqnarray} \label{328}
\alpha = 2\mu_0 -\frac{2 \mu_0}{G M_{\infty}} b + 2R \left( \frac{E}{L} \right)^2 b.
\end{eqnarray}
Then the deflection angle of light line is
\begin{eqnarray} \label{329}
\Delta \theta = 2\alpha = \frac{4G M_{\infty}}{R} -\frac{4 b}{R} + 4R \left( \frac{E}{L} \right)^2 b.
\end{eqnarray}
In \eqref{328}, the first term in the second equal sign, $\frac{4G M_{\infty}}{R}$, is the result of GR. The other two terms, which are from 4D DFT, are the minor corrections to GR for optical deflection.

\subsection{Planet precession in 4D DFT}
In this case, the particle is massive, then we introduce proper time $\tau$ and define 4-momentum of a particle:
\begin{eqnarray} \label{330}
p^{\mu} = \frac{dx^{\mu}}{d\tau}.
\end{eqnarray}
For a massive particle, we have
\begin{eqnarray} \label{331}
g_{\mu \nu } p^{\mu } p^{\nu  }  =  -1,
\end{eqnarray}
namely,
\begin{eqnarray}\label{332}
g_{00} \left( \frac{dt}{d\tau } \right)^2+ g_{11} \left(\frac{dr}{d\tau } \right)^2+g_{22} \left(\frac{d\theta }{d\tau } \right)^2+g_{33} \left(\frac{d\varphi }{d\tau } \right)^2 & = & -1.
\end{eqnarray}
The orbit of massive particle is still limited to the equatorial plane, namely, $\theta = \frac{\pi}{2}$ and $\frac{d\theta}{d\lambda}=0$.\\
For the metric \eqref{Emetric}, $p_0$ and $p_3$ are conserved, we can still get
\begin{eqnarray} \label{333}
g_{00}\frac{dt}{d\tau }  = - E,
\end{eqnarray}
and
\begin{eqnarray} \label{334}
g_{33}\frac{d\varphi }{d\tau } =  L,
\end{eqnarray}
where $E$ and $L$ represent energy and angular momentum of unit mass, respectively. Then we can obtain that
\begin{eqnarray} \label{335}
\left(\frac{dr}{d\varphi} \right)^2 = -\frac{g^2_{33}}{g_{11}}\frac{1}{L^2} - \frac{g_{33}}{g_{11}} - \frac{g^2_{33}}{g_{00} g_{11}} \frac{E^2}{L^2},
\end{eqnarray}
and
\begin{eqnarray} \label{336}
-\frac{g^2_{33}}{g_{11}} = - \frac{(r-\alpha)^2(r+\beta)^2}{\left(\frac{r-\alpha}{r+\beta} \right)^{\frac{a}{\sqrt{a^2+b^2}}}}.
\end{eqnarray}
Considering eqs.\eqref{318}, \eqref{319}, \eqref{335} and \eqref{336}, if we take $a= 2G M_{\infty}$, we can obtain the following equation about $b=0$:
\begin{eqnarray} \label{337}
\begin{aligned}
\left(\frac{dr}{d\varphi} \right)^2 & = (-r^4 + 2G M_{\infty} r^3) \frac{1}{L^2} + \frac{E^2 r^4}{L^2} -r^2 + 2G M_{\infty} r \\
   & + (6GM_{\infty} r^2 -4r^3) \frac{b}{L^2} + (2GM_{\infty}-2r)b + \frac{4E^2 r^3}{L^2}b.
\end{aligned}
\end{eqnarray}
If we define $\mu = \frac{G M_{\infty}}{r}$, then \eqref{337} can be written as
\begin{eqnarray} \label{338}
\left(\frac{d^2 \mu}{d\varphi^2} \right)^2 + \mu = \frac{G^2 M^2_{\infty}}{L^2} +3\mu^2 + \frac{6GM_{\infty}}{L^2} \mu b + \left( \frac{2GME^2}{L^2} - \frac{2GM}{L^2}\right)b,
\end{eqnarray}
where we have neglected higher orders since $\mu$ and $b$ are very small. The first two terms on the right hand side of \eqref{338} are from General Relativity, while the third term and fourth term are the contributions of $4D$ Double Field Theory (DFT).\\
We consider the nearly circular orbit whose eccentricity $e \ll 1$. The zeroth order solution of \eqref{338} is
\begin{eqnarray} \label{339}
\mu_0 = \left(\frac{G M_{\infty}}{L} \right)^2 (1 + e \cos{\varphi}),
\end{eqnarray}
then we have
\begin{eqnarray} \label{340}
3 \mu^2_0 = 3 \left(\frac{G M_{\infty}}{L} \right)^4 (1 + 2 e \cos{\varphi}),
\end{eqnarray}
where the higher order term has been neglected and
\begin{eqnarray} \label{341}
\frac{6GM_{\infty}}{L^2} \mu_0 b = 6 \frac{G^3 M^3_{\infty}}{L^4} b + 6 \frac{G^3 M^3_{\infty}}{L^4} b e \cos{\varphi}.
\end{eqnarray}
As a result, we can obtain that
\begin{eqnarray} \label{342}
\left(\frac{d^2 \mu}{d\varphi^2} \right)^2 + \mu = \left(\frac{G M_{\infty}}{L} \right)^2 + 6 \left(\frac{GM_{\infty}}{L} \right)^4 \left(1+ \frac{b}{GM_{\infty}} \right) e \cos{\varphi},
\end{eqnarray}
where the higher order constant terms have been neglected like in GR since they are the minor corrections to circular orbit, but we cannot measure the effects very accurately \cite{WeinbergGC}.\\
Let $\mu = \mu_1 + \mu_2$, where $\mu_1$ satisfies
\begin{eqnarray} \label{343}
\left(\frac{d^2 \mu_1}{d\varphi^2} \right)^2 + \mu_1 = \left(\frac{G M_{\infty}}{L} \right)^2,
\end{eqnarray}
and $\mu_2$ satisfies
\begin{eqnarray} \label{344}
\left(\frac{d^2 \mu_2}{d\varphi^2} \right)^2 + \mu_2 = 6 \left(\frac{GM_{\infty}}{L} \right)^4 \left(1+ \frac{b}{GM_{\infty}} \right) e \cos{\varphi}.
\end{eqnarray}
Eq.\eqref{343} has the general solution:
\begin{eqnarray} \label{345}
\mu_1 = \left(\frac{G M_{\infty}}{L} \right)^2 (1 + e \cos{\varphi}).
\end{eqnarray}
Eq.\eqref{344} has a special solution:
\begin{eqnarray} \label{346}
\mu_2 =3 \left(\frac{G M_{\infty}}{L} \right)^4 \left(1 + \frac{b}{GM_{\infty}} \right) e \varphi \sin{\varphi}.
\end{eqnarray}
Then the general solution of \eqref{342} is 
\begin{eqnarray} \label{347}
\begin{aligned}
\mu & = \mu_1 +\mu_2 = \left(\frac{G M_{\infty}}{L} \right)^2 (1 + e \cos{\varphi}) + 3 \left(\frac{G M_{\infty}}{L} \right)^4 \left(1 + \frac{b}{GM_{\infty}} \right) e \varphi \sin{\varphi} \\
 & \approx \left(\frac{G M_{\infty}}{L} \right)^2 \left(1+ e\cos{[1- 3 \left(\frac{GM_{\infty}}{L} \right)^2} \left(1 + \frac{b}{GM_{\infty}} \right) ] \varphi  \right),
\end{aligned}
\end{eqnarray}
where the higher order terms have been neglected.\\
The symbol of perihelion is
\begin{eqnarray} \label{348}
\left[1- 3 \left(\frac{GM_{\infty}}{L} \right)^2 \left(1 + \frac{b}{GM_{\infty}} \right) \right] \varphi = 2n\pi, 
\end{eqnarray}
where $n=0,1,2,..$. Then we have
\begin{eqnarray} \label{349}
\varphi = \frac{2n\pi}{\left[1- 3 \left(\frac{GM_{\infty}}{L} \right)^2 \left(1 + \frac{b}{GM_{\infty}} \right) \right]} \approx 2n\pi \left[1+ 3 \left(\frac{GM_{\infty}}{L} \right)^2 \left(1 + \frac{b}{GM_{\infty}} \right) \right]. 
\end{eqnarray}
The difference between the azimuth angles of two adjacent perihelion is 
\begin{eqnarray} \label{350}
\Delta \varphi =  6\pi \left(\frac{GM_{\infty}}{L} \right)^2 + 6 \left(\frac{GM_{\infty}}{L^2} \right)b. 
\end{eqnarray}
The first contribution of \eqref{350} is from General Relativity, while the second term is from 4D DFT. 

\subsection{Gravitational redshift in 4D DFT}
When a photon is propagating in static gravitational field, stationary observers in different locations will observe different frequencies, which is called gravitaional redshift of photons \cite{WeinbergGC}. For static gravitational field, we can choose coordinate properly, which makes $g_{\mu\nu}$ does not dependent on $t$. Then for a stationary observer, the observed energy of photon is given by:
\begin{equation} \label{351}
E=-p_{\mu} U^{\mu},
\end{equation}
where $p_{\mu}$ is the 4-momentum of photon and $U^{\mu}$ is the 4-velocity of the observer, which has the following form:
\begin{equation} \label{352}
U^{\mu} = \frac{1}{\sqrt{-g_{00}}} (1,0,0,0).
\end{equation}
Combining \eqref{351} and \eqref{352}, we have
\begin{equation} \label{353}
E = \frac{-p_0}{\sqrt{-g_{00}}}.
\end{equation}
According to the Planck formula $E=h\nu$, then \eqref{353} can be written as
\begin{equation} \label{354}
\sqrt{-g_{00}} \nu = -p_0 / h.
\end{equation}
The right hand side of \eqref{354} is a constant, since $p_0$ is a constant when a particle is moving in static gravitational field. Therefore, the frequency of photon is inversely proportional to $\sqrt{-g_{00}}$ locally. \\
According to \eqref{Emetric} and $b \ll a$, then in the Einstein frame we have
\begin{equation} \label{355}
-g_{00} \approx 1-\frac{a}{r} + \frac{(a-2r)(1+\frac{a}{r})}{(a-r)r}b.
\end{equation}
If we set $a=2G M_{\infty}$, then 
\begin{equation} \label{356}
-g_{00} \approx 1-\frac{2G M_{\infty}}{r} + \frac{(2G M_{\infty}-2r)(1+\frac{2G M_{\infty}}{r})}{(2G M_{\infty}-r)r}b,
\end{equation}
and
\begin{equation} \label{357}
\sqrt{-g_{00}} \approx 1-\frac{G M_{\infty}}{r} + \frac{(G M_{\infty}-r)(1+\frac{2G M_{\infty}}{r})}{(2G M_{\infty}-r)r}b.
\end{equation}
In weak gravitational field, the redshift is small, then we obtain that
\begin{equation} \label{358}
\frac{\nu}{\nu_0} = 1-\frac{G M_{\infty}}{R} + \frac{(G M_{\infty}-R)(1+\frac{2G M_{\infty}}{R})}{(2G M_{\infty}-R)R}b,
\end{equation}
where $\nu$ is the frequency of photon in infinity, $\nu_0$ is the frequency of photon on the surface of star, and $R$ is the radius of star. \\
We often define redshift $z = \frac{\nu_0}{\nu}-1$, then we have
\begin{equation} \label{359}
z = \frac{G M_{\infty}}{R} - \frac{(G M_{\infty}-R)(1+\frac{2G M_{\infty}}{R})}{(2G M_{\infty}-R)R}b.
\end{equation}
Then second term of \eqref{359} is the correction to gravitaional redshift in Gerneral Relativity.

\section{Conclusions and Discussions}
Although General Relativity (GR) is a very successful theory of gravity, it still cannot explain every observational phenomenon, such as the galaxy rotation curve. The resolution of the discrepancy might call for dark matter or modifications of the theory of gravity, or perhaps both \cite{KPS2017,M1983}. At present the most promising quantum gravity theory is string theory, which is believed to unify four fundamental interactions. Therefore, it is natural to test string theory to explain the phenomena which GR cannot explain. \\
In recent years, Double Field Theory (DFT) has been an exciting research area in string theory. DFT intends to reformulate supergravity with doubled coordinates, namely, $x_A = (\tilde{x}_{\mu}, x^{\nu})$ \cite{KPS2017, Siegel19931, Siegel19932, HZ20091}. As a result, DFT unifies diffeomorphism and B-field gauge symmetry into ‘doubled diffeomorphisms’ and T-duality has been a manifest symmetry of the action of DFT \cite{HZ20092,HZ20101,HZ20102}. The most general, spherically symmetric, asymptotically flat, static vacuum solution to $D=4$ DFT has been obtained in \cite{KPS2017}. Stephen Angus, Kyoungho Cho and Jeong-Hyuck Park studied more properties of Einstein double field equations further \cite{ACP2018}. Choi and Park performed post-Newtonian analysis of double field theory as a test of string theory in gravitational sector against observations in \cite{CP2022}. Yang Liu analyzed Hawking radiation of the solution \cite{YL2022}, which was obtained in \cite{KPS2017}. \\ 
In this article we calculate the minor corrections to the three predictions in General Relativity: optical deflation, planet precession and gravitational redshift, namely, \eqref{329}, \eqref{350} and \eqref{359}. In recent years more and more cosmological observations have been conducted. One of the most significant developmets is about supermassive black holes \cite{FM2000, V2010, IVH2020}. Hence, the value of $b$ should be determined by observational data. The results we obtained in this article, should be able to be tested accurately as well. We hope that people could find suitable star or black hole systems to test our results in the future. \\
Future work can be directed along at least three lines of further research. First, the metric obtained in \cite{KPS2017} should be generalized to other cases, such as rotating black holes or charged black holes. Second, 4D Double Field Theory should be applied to explain cosmological phenomena, such as dark matter and dark energy. Third, Exceptional Field Theory (ExFT) should be applied to black hole and cosmology. There is therefore great potential for development of this work in the future.

\paragraph{Acknowledgements} 
We would like to thank Zhiyang Bao and Junhan Liu for useful discussions. YL was supported by an STFC studentship. For the purpose of open access, the authors have applied a CC BY public copyright licence to any Author Accepted Manuscript version arising.

\appendix

\end{document}